\documentclass[aps,12pt,reprint,amsmath,amssymb,amsfonts,aps,prb,twocolumn,floatfix,nofootinbib,showpacs,longbibliography, includegraphics]{revtex4-2}

\usepackage{mathtools}
\usepackage{braket}
\usepackage{bm}
\usepackage{booktabs}
\usepackage{multirow}
\usepackage[indent=15pt]{parskip}
\usepackage{graphicx} 
\usepackage{libertinus}  
\usepackage{libertinust1math}
\usepackage[T1]{fontenc}
\usepackage[colorlinks, allcolors=blue]{hyperref}
\usepackage[dvipsnames]{xcolor}

\begin{abstract}
It is believed that an isolated and far-from-equilibrium quantum many-body system should try to attain equilibrium via a mechanism whereby any given subsystem acts as an open quantum system that is coupled to an environment, which is the complementary part of the full system, and undergoes a complicated equilibration process such that all the subsystems in the long-time limit attain equilibrium states compatible with the global equilibrium state. This picture begs the question whether the dynamics of any given subsystem is Markovian (monotonic loss of information and memory) or non-Markovian. In this work, by numerically probing the dynamical behaviour of the quantum distances between $\textit{temporally-separated}$ states of small subsystems, we reveal the telltale signatures of (non-)Markovianity of the dynamics of subsystems of an isolated quantum spin system brought in the far-from-equilibrium regime, exemplified with the mixed-field Ising spin chain quenched between parameter regimes deep inside its magnetically ordered and disordered regimes. Additionally, remarkably systematic behaviour is seen in a measure of classical distances between the quantum states of the considered subsystems. These features strongly depend on the direction of quenching in the parameter space, with paramagnetic-to-ferromagnetic quenches offering considerably stronger signatures of subsystem non-Markovianity, for which we offer heuristic arguments.
\end{abstract}

\begin{document}

\title{Non-Markovianity of subsystem dynamics in isolated quantum many-body systems}

\author{Aditya Banerjee}
\email{adityabphyiitk@gmail.com}
\affiliation{Theory Division, Saha Institute of Nuclear Physics, 1/AF
Bidhannagar, Kolkata 700064, India}


\maketitle

\section{Introduction}   \label{sec1}

It is expected that a closed quantum many-body system when brought far out of equilibrium by quenching one of its parameters will, after a sufficiently long-time, equilibriate to a state as dictated by the post-quench parameters. The actual process of equilibration can be rather complex because during this process the system will explore the set of accessible states (decided by the extant conserved quantities) in the Hilbert space during its Hamiltonian dynamics, and the complexity of the trajectories in the quantum state space depends crucially on the initial state and its relation with the Hamiltonian driving its evolution. For generic (i.e., non-integrable) systems with only the total energy being a conserved quantity, the long-time physics is described by usual statistical mechanics in the microcanonical picture, which essentially encapsulates the notion of ergodicity. One then says that the system has thermalized, i.e., the long-time (thermal) state is the Gibbs density matrix as described by the Gibbs ensemble, and the eigenstate thermalization hypothesis (ETH) describes well the behaviour of observables in this regime \cite{Alessio2016,Gogolin2016,Mori2018,Deutsch2018}. Exceptions to this paradigm within the class of non-integrable quantum many-body systems exist of course \cite{Abanin2019,Nandkishore2015,Serbyn2021,Chandran2023}. For integrable quantum many-body systems, i.e., those with an extensive number of conserved quantities, the long-time physics is instead described by a so-called generalized Gibbs ensemble picture wherein the equilibrated state is a generalized Gibbs density matrix \cite{Essler2016,Vidmar2016}, and one says that these systems do not thermalize in the usual sense (but possibly in a generalized sense \cite{Cassidy2011}) and constitute a strong violation of ergodicity and the ETH.  

However, the nature and characteristics of the Hamiltonian dynamics in the far-from-equilibrium regime (early and intermediate times), where organizing principles or structures are scarce, is much less understood. In particular, it seems prudent to have a broader classification of general characteristics of the dynamics that goes beyond the existing classifications such as fast-, slow- or non-thermalizing/equilibrating as measured by e.g. the growth of entanglement entropies or the behaviour of local observables with time. One natural way to think about this dynamical process is by considering any subsystem of the full system as an \textit{open} quantum system interacting with its environment, i.e., its complementary part of the full system, and it is natural to presuppose that it is in fact the nature of the interactions between the various subsystems acting as environments for each other that decides the various facets of the non-equilibrium relaxation dynamics.

With this line of thinking, and employing the notion of distances or divergences between quantum states of which many measures exist in (quantum) information theory \cite{Wilde2017,Hayashi2017,Watrous2018}, this work investigates the following basic question : how does the distance between \textit{temporally separated} quantum states of a given \textit{subsystem} evolve in time and what underlying characteristics of the non-equilibrium dynamics are thereby revealed. Independently, the viewpoint of regarding a given subsystem as an open quantum system necessitates asking whether its dynamics is Markovian (memory-less, and monotonic loss of information) or not \cite{Rivas2014,Breuer2016,Li2018,Chruscinski2022,Shrikant2023}. Both these questions are intimately related to each other, and we believe this classification of subsystem dynamics as Markovian or non-Markovian is a concrete addition to the existing list of characteristics of the non-equilibrium dynamics of closed quantum many-body systems (see also \cite{Romero2019}). 

In this work we have focused on investigating small subsystems comprising of a few spins embedded in the (translationally-invariant) bulk of a large spin chain. This is partly due to the fact that the computational costs of constructing reduced density matrices (RDM) scale exponentially in the subsystem sizes, and in fact already for subsystems of only five spins this was evident to us in practice. This also partly served the purpose of going systematically in investigating the dynamics of subsystems starting up from the smallest possible one. We will see later that already for a four-spin subsystem, the non-Markovianity of its dynamics is quite insignificant, so our focus on small subsystems gets incentivized \textit{a posteriori}. Perhaps most importantly, in contemporary ultracold atomic experiments, local site-resolved studies of quantum states has been made feasible and put to practice (see e.g. \cite{Ott2016,Kuhr2016,Gross2021}), which motivates putting the microscope on very small subsystems also theoretically. In fact, given that experimental access and control is often limited to small subsystems and not the whole of large quantum many-body systems, probing and understanding the various attributes of the dynamics of small subsystems (as experimentally accessible open systems surrounded by the inaccessible environments) is thus also a pertinent theoretical task.

This article is organized as follows. We introduce our notations and some preliminary notions in Section \ref{sec2}, Section \ref{sec3} introduces and discusses the concept and a common measure of information backflow, and Markovianity and its lack thereof, as well as how to quantify them in the context of isolated quantum many-body dynamics. In Section \ref{sec4}, we present our results for the mixed-field Ising spin chain, using the trace distance measure (defined in Eq.(\ref{TD})) as our figure of merit, with the main result being that the paramagnetic-to-ferromagnetic quenches induce significant information backflow and non-Markovianity of subsystem dynamics, while the oppositely-directed quenches exhibit essentially Markovian dynamics of small subsystems. In addition, notably systematic behaviour in the dynamical behaviour of a corresponding "classical" distance measure (defined in Eq.(\ref{TVD})) is also revealed and discussed. Section \ref{sec5} provides an extended discussion of several aspects of our work and related matters, and Section \ref{sec6} provides concluding remarks. 

\section{Preliminaries}  \label{sec2}

\textit{Quantum states and channels---}  Herein we introduce our notions and notations. Quantum states live in the Hilbert space $\mathcal{H}$, which for quantum spin systems on $N$ sites assumes a tensor product structure $(\mathbb{C}^d)^{\otimes N}$ over the $N$ local Hilbert spaces each of dimension $d$ (e.g. $d=2$ for $S=1/2$ spins). Often though by quantum states we shall refer to the density matrix (operator) $\rho$. It is Hermitian and positive semi-definite with unit trace, which means that its eigenvalues $\{ p_i\} $ can be regarded as providing a bonafide probability distribution. The set of density matrices is convex and will be denoted by $ \mathcal{D}(\mathcal{H})$.

A trace-preserving quantum operation $\Lambda$ is called positive if $\Lambda \! : \! \mathcal{D}(\mathcal{H}) \! \rightarrow \! \mathcal{D}(\mathcal{H})$ (this holds more generally over the space of linear and bounded Hermitian operators, of which $\mathcal{D}(\mathcal{H})$ is a particular subset). A more powerful and non-trivial notion is complete-positivity, which means that $\mathbb{I} \otimes \Lambda$ is positive, where the identity operation $\mathbb{I}$ is understood to act on those parts of the system that are complementary or ancillary to the support of the action of $\Lambda$. Completely-positive trace-preserving (CPTP) maps are also known as quantum channels. Examples include unitaries, measurements and partial trace. Physically (meaning, experimentally) realizable operations have to be CPTP.

\textit{Distances between quantum states---} We choose the trace distance as our primary measure of distance between quantum states in this article (results are qualitatively unchanged if other measures are chosen). Our choice for this measure is motivated by simplicity and ease of computability, as the trace distance measure between quantum states is free from any potential numerical issues that may arise with other measures involving fractional powers or logarithms of sufficiently large density matrix operators. Given two density matrices $\rho$ and $\sigma$, the trace distance (TD) between them is defined as,
\begin{equation}     \label{TD}
    T_d(\rho,\sigma) = \frac{1}{2}\sum|\lambda_i|  \text{    ,}
\end{equation}
where $\{\lambda_i\}$ is the set of eigenvalues of $(\rho-\sigma)$, with the index $i\!=\!1,2,..,\operatorname{rank}(\rho-\sigma)$ . The trace distance takes values $\in [0,1]$. Let $\{p^{\downarrow}\}$ and $\{q^{\downarrow}\}$ be the vector of eigenvalues of $\rho$ and $\sigma$ respectively, arranged in a descending order. The counterpart of trace distance for probability distributions, known as the total variation distance (TVD), is defined as, 
\begin{equation}     \label{TVD}
    V_d(p^{\downarrow},q^{\downarrow})=\frac{1}{2}\sum_i |p^{\downarrow}_i - q^{\downarrow}_i|  \text{   ,}
\end{equation}
and it also takes values $\in [0,1]$. This can be thought of as the classical counterpart to the trace distance, and encodes in some sense a notion of classical distances between the respective density matrices.

\section{Assessing (non-)Markovianity}   \label{sec3}

A fundamental and powerful property of any distance measures is that they are non-increasing (or "contractive") under CPTP maps \cite{Wilde2017,Hayashi2017,Watrous2018,Holevo2019}. That is, given a CPTP operation $\Lambda:\mathcal{D(\mathcal{H})} \! \rightarrow  \! \mathcal{D(\mathcal{H})}$, one has, for the trace distance,
\begin{equation}      \label{TDCPTP}
    T_d(\Lambda(\rho),\Lambda(\sigma)) \leq T_d(\rho,\sigma)    \text{   .}
\end{equation}
In other words, any (physically realizable) quantum operation can not increase the (information about the) distance between (and hence, a quantifier of "distinguishability") of two quantum states.

Consider now a dynamically evolving quantum system. Let $\Lambda_{t,0}$ denote the family of CPTP maps that evolves the subsystem from initial time $t=0$ to time $t$ for all $t>0$, these will be referred to as dynamical maps. This temporal evolution process is said to be (CP-)\textit{divisible} if one has,
\begin{equation}        \label{CPdiv}
    \Lambda_{t,0} = \Lambda_{t,s}\Lambda_{s,0} \hspace{0.3cm} \forall s \in (0,t)  \text{    ,}
\end{equation}
where $\Lambda_{t,s}$ is also a CPTP map evolving the system from time $s$ to $t$, otherwise if $\Lambda_{t,s}$ is not CPTP, the evolution is termed (CP-)indivisible (note that it is not to be taken for granted nor is it obvious or generic that the intermediate map $\Lambda_{t,s}$ has to be CPTP) \cite{Wolf2008a}. It then follows that in a divisible process, the aforementioned monotonicities are respected at every step of the process, and so ultimately one has,
\begin{equation}       \label{TDmonotone}
    T_d(\Lambda_{t,0}(\rho),\Lambda_{t,0}(\sigma)) \leq T_d(\rho,\sigma)  \text{    ,}
\end{equation}
 for any $t > 0$. The distance between two quantum states continues to decrease all along the temporal dynamics, and such a dynamical process is termed \textit{Markovian}, and any violation of this condition can be taken as a signature of \textit{non-}Markovianity \cite{Rivas2010,Rivas2014,Breuer2016}. Thus, an CP-indivisible process is non-Markovian though the converse does not necessarily hold true \cite{Milz2019}. An alternative approach is to view the upholding of these monotonicities as reflecting the monotonic loss of information ("memory") about the distinguishable features of a pair of quantum systems coupled separately to the \textit{same} environment, which is stable by itself and is not affected by its interactions with the non-equilibrium quantum systems to which it is coupled (which is the essence of the Born-Markov approach to open quantum systems \cite{Breuer2002,Vacchini2024}). Such an environment acts to drive the systems to the final equilibrium states and thereby decreasing any distinguishable features (absorbing the information about such distinguishabilities) between them over time; such a process is Markovian by this token \cite{Breuer2016}. Consequently, any violation of these monotonicities signal backflows of information from the environment to the systems (i.e., the environment somehow retained the dynamical memory of the past states of the systems; the system-environment coupling is beyond the Born-Markov regime and the environment is in fact itself affected by the systems it is coupled to and unable to quickly shed information about them) and this is a defining signature of non-Markovianity \cite{Breuer2009,Breuer2010}. These two viewpoints are not necessarily equivalent to each other in general nor is it obvious that they should be, however for invertible CPTP maps (that is, $\Lambda_{a,b}^{-1}$ exists, meaning none of the eigenvalues of $\Lambda_{a,b}$ is zero; such is the case for us as explicitly checked numerically) these two viewpoints are understood to be equivalent \cite{Bylicka2017,Chruscinski2018,Chakraborty2018,Chruscinski2022}. It is important to remark here that invertibility is not to be confused with reversibility, and invertible maps are not generally CPTP either. Only for a unitary CPTP map is its inverse also CPTP and consequently invertibility is the same as reversibility, see e.g. \cite{Chruscinski2022} for extended discussions on this subtle point. Moreover, for certain classes of dynamical maps, invertibility has been argued to be necessary for Markovianity \cite{Dugic2023}, though clearly it can not be sufficient. To be sure, our numerical results are really only measuring the backflows of information, and not actually establishing if all the intermediate $\Lambda_{t,s}$ are CPTP or not and thereby establishing (in)divisibility from first principles, which is a separate issue in itself and is being addressed in a parallel work. It is only thanks to these maps being invertible that we talk of both these notions of non-Markovianity interchangeably in this work.  

Consider then a subsystem (not necessarily contiguous) of length $l$ spins, represented by a density matrix $\rho_t^{(\ell)}$ at time $t$, of a non-equilibrium quantum many-body system that is evolving under its own Hamiltonian ($t=0$ denotes the starting time). We shall be dealing exclusively with translationally invariant systems in this work, thus it does not matter where exactly this size-$\ell$ subsystem is located within the full system as long as it is located in the bulk away from the boundaries (in case of open boundary conditions). We probe the following basic questions in this work : how the distance between two \textit{temporally-separated} states of a subsystem behaves as the system evolves in time, and what characteristics of the non-equilibrium dynamics is revealed by this behaviour, and if there are any invariants or  qualitative changes under varying temporal separation of considered subsystem states ? As such, our main quantities of interest are $T_d(\rho_{t+\delta}^{\ell}, \rho_t^{\ell})$, where $\delta$ is the temporal separation. Let $\Lambda_{(t)}$ be the dynamical CPTP map that evolves the subsystem by time $t$, i.e., 
\begin{equation}    \label{map}
    \rho_{t+a}^{\ell} = \Lambda_{(t)}[\rho_{a}^{\ell}]  \text{   .}
\end{equation}
Then by previous discussion, if the dynamics is divisible, i.e., $\Lambda_{(t)} \!=\! (\Lambda_{(1)})^{t}$ (more generally, in terms of the previous discussion, $\Lambda_{(t)}\!=\!\Lambda_{t+a,a}$ for any reference time stamp $a$), then a non-increasing behaviour of $T_d(\rho_{t+\delta}^{\ell}, \rho_t^{\ell})$ and consequently no information backflow must be observed for \textit{any} $\delta$. Consequently, any violation of this monotonically non-increasing behaviour of these two quantities is a signature of indivisibility as well as a backflow of information (about the distinguishability between quantum states) from the environment back to the subsystem in question and thus, of non-Markovianity, and a degree of non-Markovianity may then be defined by the cumulative magnitude of such violations. Speaking in terms of discrete times as in numerical simulations (continuous-time generalization is obvious) with a time-step $\tau$, let the discrete "slope" of the aforementioned trace distance for a given $\delta$ be denoted as, 
\begin{equation}    \label{slope}
    \alpha(t,\delta) = \frac{1}{\tau}\bigg(T_d(\rho_{t+\tau+\delta}^{\ell}, \rho_{t+\tau}^{\ell}) - T_d(\rho_{t+\delta}^{\ell}, \rho_{t}^{\ell}) \bigg)   \text{    .}
\end{equation}
This quantity is always negative for Markovian dynamics. Thus, a degree of non-Markovianity measuring the cumulative magnitude of revivals (increases) of the trace distance $T_d(\rho_{t+\delta}^{\ell}, \rho_t^{\ell})$ can be defined as,
\begin{equation}     \label{TDdegree}
    \mathcal{D}(\delta) = \sum_t \alpha(t,\delta)  \hspace{0.3cm} \forall t \hspace{0.25cm} \text{s.t.} \hspace{0.25cm} \alpha(t,\delta)>0    \text{ .}
\end{equation}
Note that this differs from the degree of non-Markovianity used in \cite{Breuer2009,Breuer2010} where a maximization over pairs of initial states has been made due to their interests in defining this degree for a given quantum dynamical process acting on various initial states. Our interest however is slightly different, in that we are concerned with a fixed initial state for a given class of quenching and wish to quantify the non-Markovianity arising out of quenching it by parameters lying in a different phase of the underlying system, and then comparing this degree for different quenching parameters within the same post-quench parameter regime. That is, for example, in paramagnetic-to-ferromagnetic class of quenches, the initial paramagnetic phase is fixed and a non-Markovianity degree is calculated for each ferromagnetic quenching parameters, and then these degrees are compared with each other for different ferromagnetic quenching parameters. Later in Sec.\ref{sec4}, we shall also consider the dynamical behaviour of the trace variation distance Eq.(\ref{TVD}) between the descendingly-ordered eigenvalues of the temporally separated subsystem density matrices, and a quantity measuring the cumulative magnitude of revivals of the TVD measure can be similarly defined. Let the corresponding slope for the TVD measure for a given $\delta$ be, 

\begin{equation}    \label{slope1}
    \alpha_1(t,\delta) = \frac{1}{\tau}\bigg(V_d(q_{t+\tau+\delta}^{\ell}, q_{t+\tau}^{\ell}) - V_d(q_{t+\delta}^{\ell}, q_{t}^{\ell}) \bigg)   \text{    ,}
\end{equation}
where $q_{t}^{\ell}$ denotes the eigenvalues, arranged in descending order, of the density matrix $\rho_t^{\ell}$. The degree of TVD revivals can then be defined as, 
\begin{equation}     \label{TVDdegree}
    \mathcal{D}_1(\delta) = \sum_t \alpha_1(t,\delta)  \hspace{0.3cm} \forall t \hspace{0.25cm} \text{s.t.} \hspace{0.25cm} \alpha_1(t,\delta)>0    \text{ .}
\end{equation}

\section{Results and discussion for the mixed-field Ising spin chain}   \label{sec4}

We demonstrate our results on the paradigmatic one-dimensional Ising spin chain in the presence of both transverse and longitudinal fields with open boundary conditions,
\begin{equation}     \label{ising}
    \mathcal{H}_I = -J\sum_{j=1}^{N-1}\sigma^z_j \sigma_{j+1}^z - h_x\sum_{j=1}^N \sigma_j^x - h_z\sum_{j=1}^N \sigma_j^z  \text{   
        .}
\end{equation}
When $h_z\!=\!0$, the model is integrable, with the Ising critical point at $J\!=\!h_x$ separating the symmetry-broken ferromagnetic/antiferromagnetic phases from the paramagnetic phase \cite{Pfeuty1970}. When $h_z\!\neq\! 0$, the model is non-integrable and in the symmetry-broken phases (ferromagnetic or anti-ferromagnetic) exhibits confinement between kink-antikink excitations \cite{McCoy1978,Delfino1996,Fonseca2003,Rutkevich2008} and consequently slow thermalization (and other features associated with it) after quenching to this non-integrable regime \cite{Banuls2011,Kormos2016,Lin2017,Alvaredo2020,Liu2019,Robinson2019,Scopa2022,Birnkammer2022,Knaute2023,Kaneko2023,Robertson2024}.

For our simulations we use the matrix product state (MPS) formalism and second-order time-evolving block decimation (TEBD2) algorithm for simulating the real-time dynamics \cite{Vidal2004,Schollwoeck2011,Paeckel2019}. Our simulations were performed with the ITensors library in Julia \cite{Itensor,ITensor-r0.3}. Most results shown here were obtained with Trotter time-steps $\tau\!=\!0.01$ for our TEBD2 simulations (thus, our numerical errors are $\mathcal{O}(10^{-4})$), and we have checked that the results were unchanged with a smaller time-step of $\tau\!=\!0.002$. The total system size was fixed at $N\!=\!200$ sites, and results were independent of the other system sizes used for verification purposes. The MPS cutoffs were set at $10^{-9}$, and maximum bond-dimensions were set at $50$. The ground states were obtained with the density matrix renormalization group (DMRG) algorithm \cite{Schollwoeck2011}, also implemented with the same cutoffs as mentioned above. The unit of time is $J^{-1}$ for paramagnetic-to-ferromagnetic quenches, and $h_x^{-1}$ for ferromagnetic-to-paramagnetic quenches.

\begin{figure}[tp]
	\centering
	\includegraphics[width=4.2cm,height=3cm]{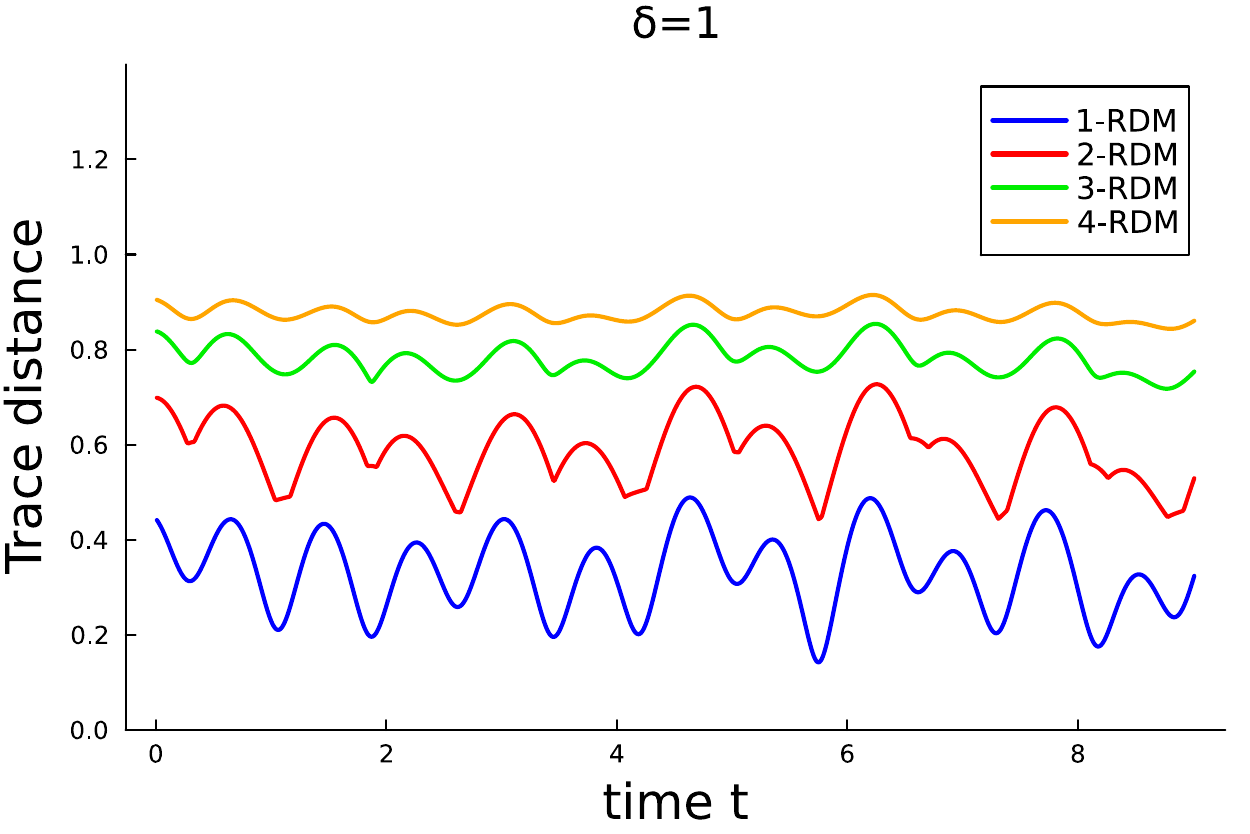}
    \includegraphics[width=4.2cm,height=3cm]{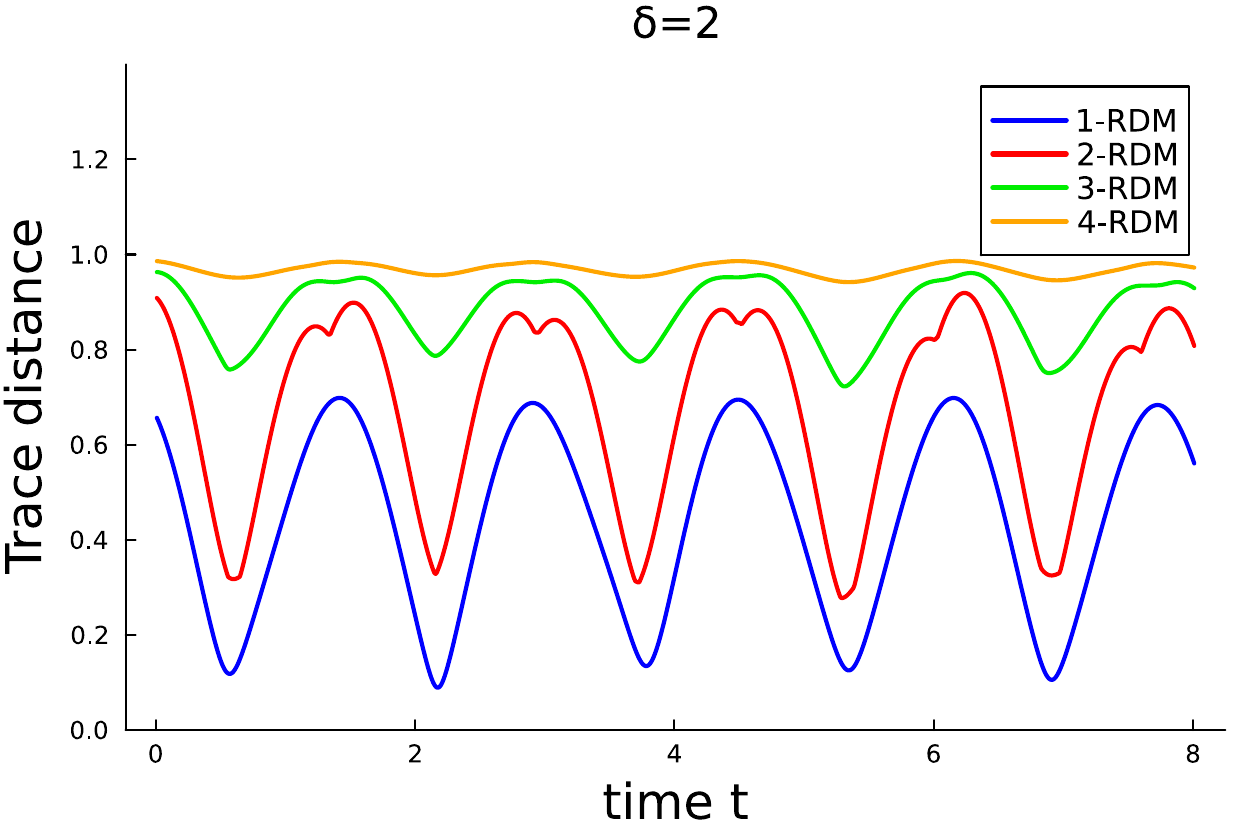}
	\caption{\fontsize{9}{11} \selectfont Paramagnetic-to-ferromagnetic quench $(J,h_x,h_z) \!=\! (0.2,1,0) \rightarrow  (1,0.1,0.5)$ : Trace distances (TD), Eq.(\ref{TD}) (between $\delta-$separated RDMs of small subsystems), plotted on the y-axes vs time for various small subsystem RDMs with $\delta\!=\!1$ (\textbf{left}), and $\delta\!=\!2$. (\textbf{right}). In the labels, $k-$RDM refers to subsystem size, $k\!=\!\{1,2,3,4\}$. The highly non-monotonic behaviour signifies strong non-Markovianity.}
	\label{fig:fig1}
\end{figure}

\begin{figure}[tp]
	\centering
	\includegraphics[width=4.2cm,height=3cm]{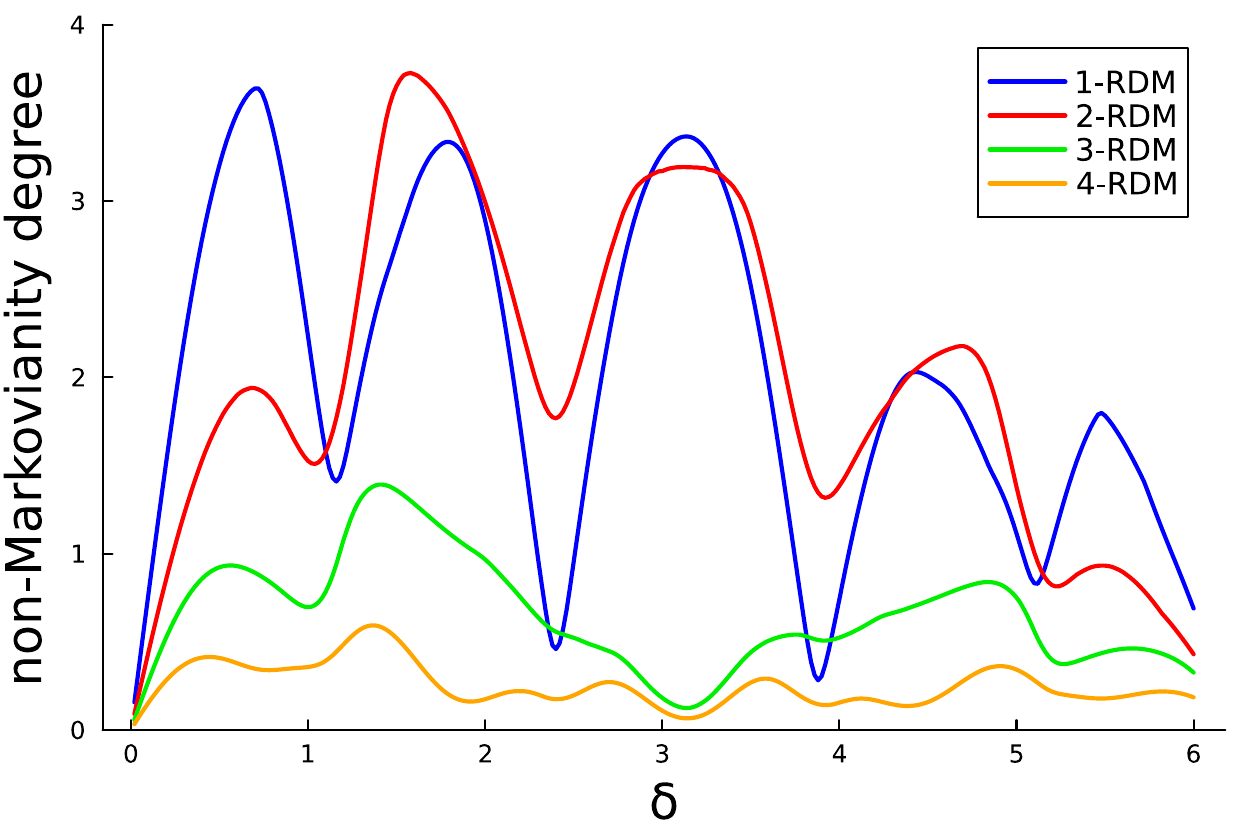}
    \includegraphics[width=4.2cm,height=3cm]{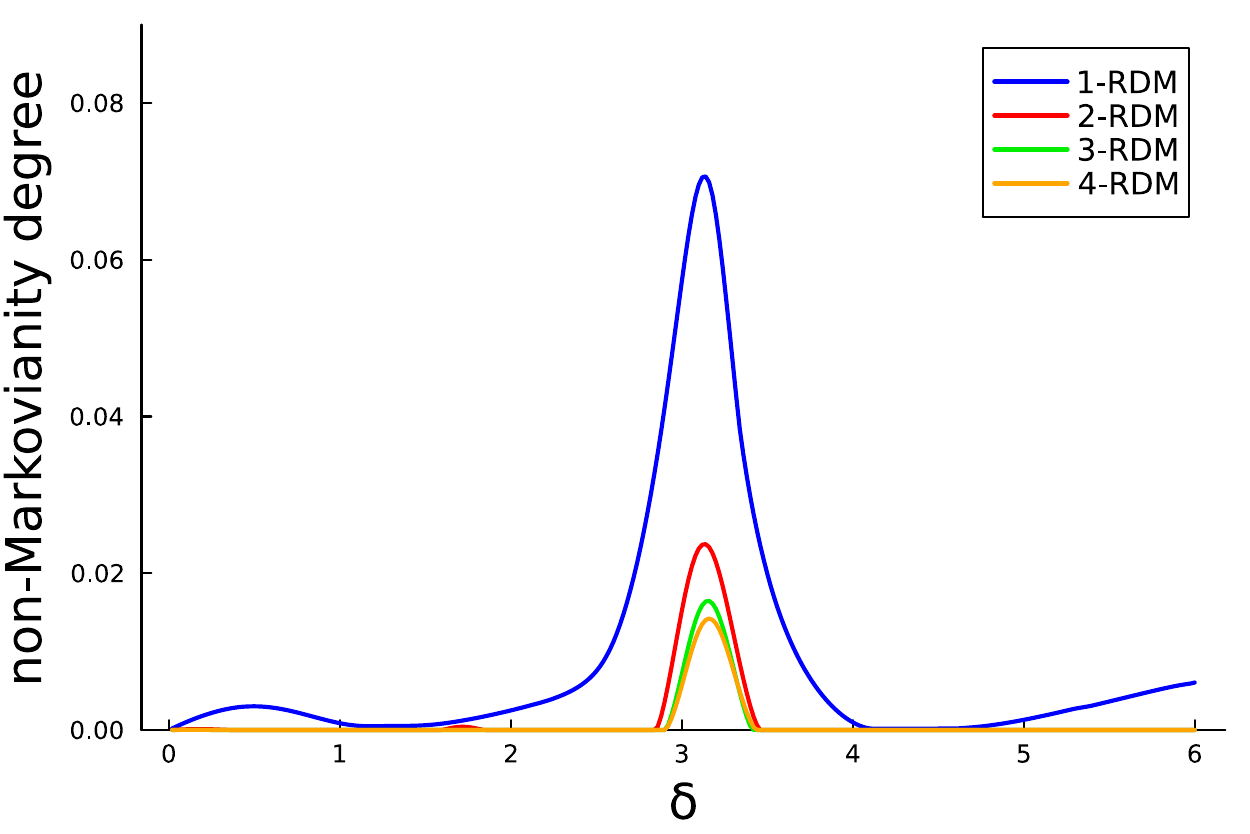}   
    \includegraphics[width=4.2cm,height=3cm]{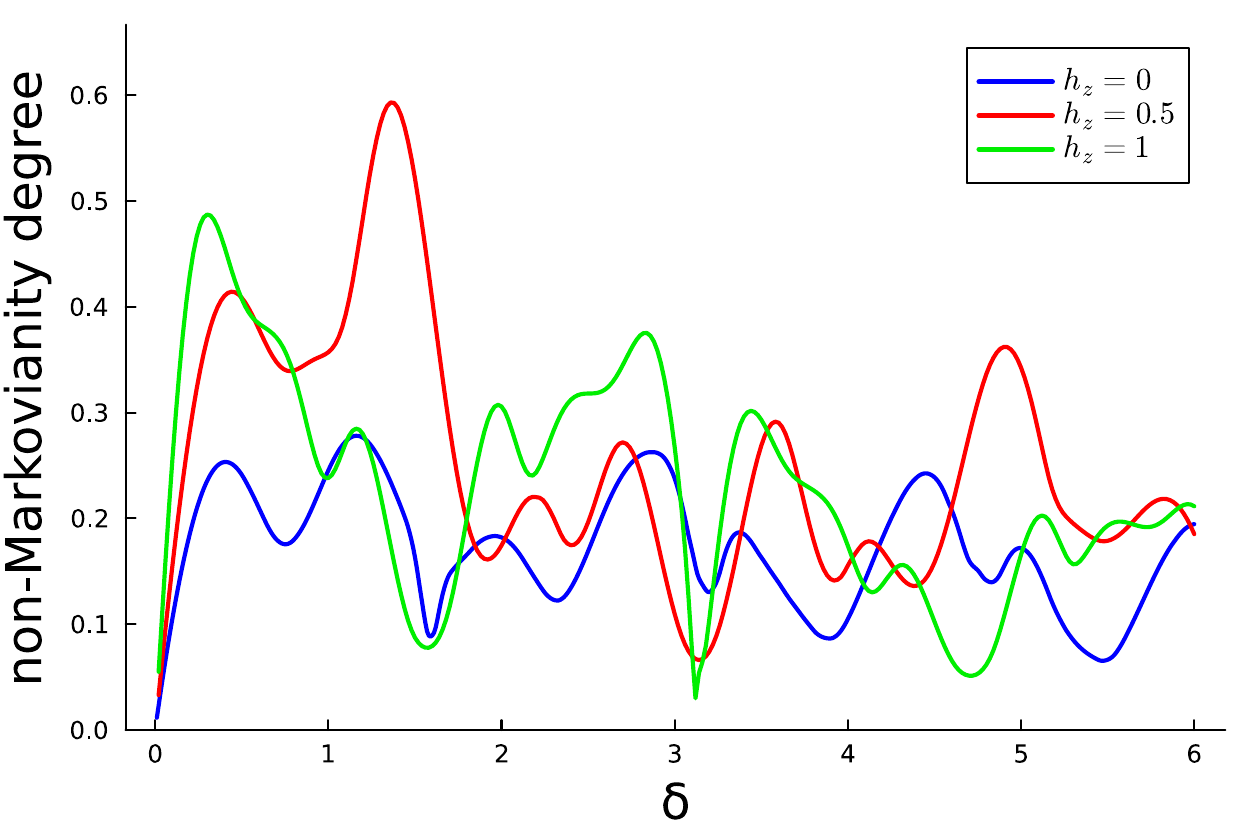}
    \includegraphics[width=4.2cm,height=3cm]{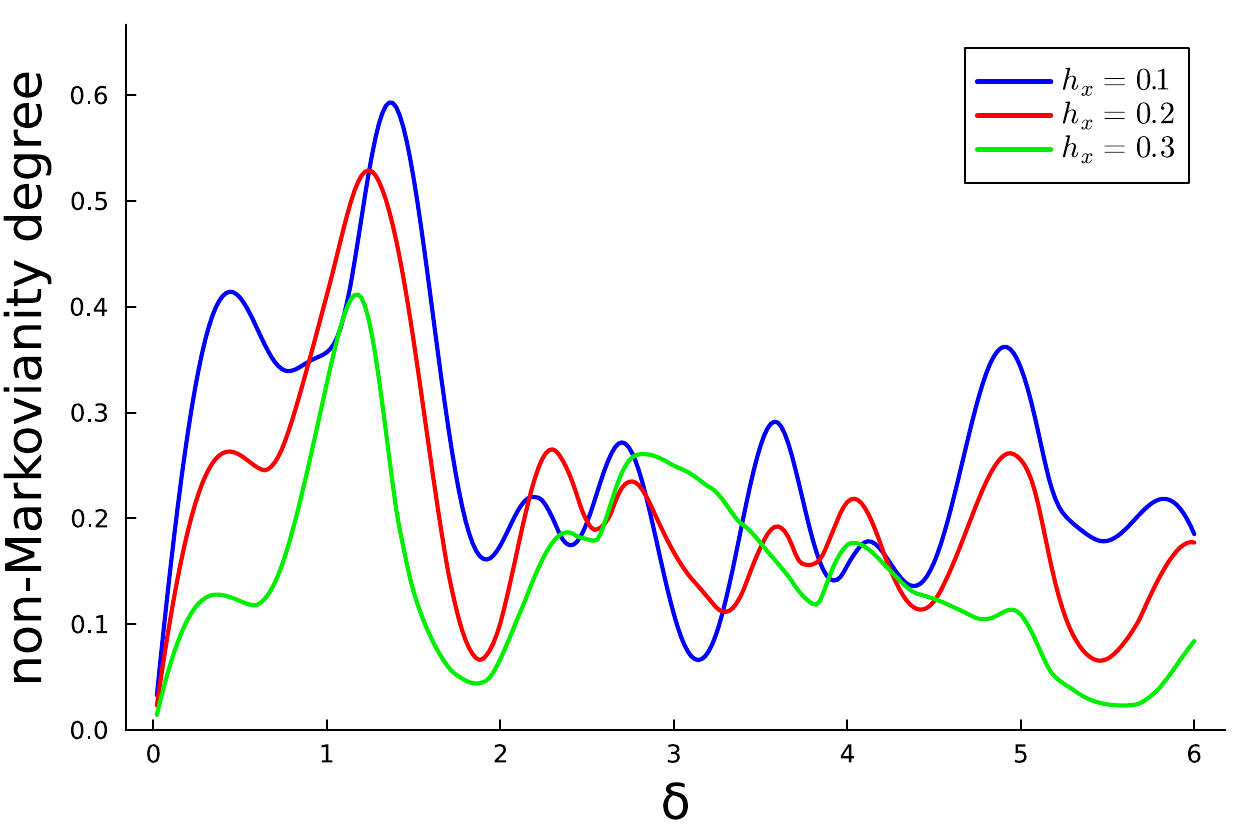}
    \caption{\fontsize{9}{11} \selectfont Degree of subsystem non-Markovianity for TD distances $\mathcal{D}(\delta)$, Eq.(\ref{TDdegree}), on the y-axes vs $\delta$. \textbf{Upper Row} - For various small subsystems in paramagnetic to ferromagnetic quench $(J,h_x,h_z) \!=\! (0.2,1,0) \rightarrow  (1,0.1,0.5)$ (left), and ferromagnetic to paramagnetic quench $(J,h_x,h_z) \!=\! (1,0.1,0.5) \rightarrow  (0.2,1,0) $ (right). Note that this degree of subsystem non-Markovianity in the latter quench are smaller than in the former quench by a factor of about 50 or more. \textbf{Lower Row} - For 4-RDMs with varying longitudinal field $h_z$ (left) and transverse field $h_x$ (right) in the case of paramagnetic-to-ferromagnetic quenches.}
        \label{fig:fig2}
\end{figure}

In Fig.\ref{fig:fig1}, we show results for the evolution of trace distance $T_d(\rho_{t+\delta}^{\ell}, \rho_t^{\ell})$, for subsystems composed of one ($\ell\!=\!1$) to four spins in contiguous blocks ($\ell\!=\!4$), separated in time by $\delta\!=\!1$ and $2$ for a quench starting from the paramagnetic ground state at $(J,h_x,h_z)\!=\!(0.2,1,0)$ to the ferromagnetic side $(J,h_x,h_z)\!=\!(1,0.1,0.5)$ (in the labels of Fig.\ref{fig:fig1}, $k-$RDM refers to reduced density matrices (RDMs) of subsystem size, $k\!=\!\{1,2,3,4\}$). Note that the latter parameters are deep in the non-integrable regime. The highly non-monotonic behaviour of the trace distances in Fig.\ref{fig:fig1} is a telltale signature of strong non-Markovianity in the dynamics of these subsystems. Moreover, the oscillatory behaviour in trace distances appear to be rather persistent. This, we believe, is also a signature of very slow relaxation and anomalous thermalization, since it stands to reason that strong non-Markovianity should in general be an obstruction to fast and efficient relaxation. Note also that at any fixed time $t$, the trace distances between the temporally-separated states of the larger subsystems are strictly higher than those of the smaller subsystems : this is simply due to the contractivity of any distance measures under the partial trace operation (a CPTP operation).

A more revealing picture may be obtained by plotting the degree of non-Markovianity Eq.(\ref{TDdegree}) vs the temporal separation $\delta$. Fig.\ref{fig:fig2} shows this degree, the upper row of which reveals two facts : I) no discernible mathematical pattern is seen for the case of paramagnetic-to-ferromagnetic quench except for the fact that this degree assumes noticeably smaller values for the larger subsystems of three and four spins, signifying that the dynamics of smaller subsystems of one and two spins are considerably more non-Markovian and further larger subsystems have progressively lessening non-Markovianity in their dynamics, and II) a negligible degree of non-Markovianity is seen in the case of ferromagnetic-to-paramagnetic quench (about $\sim 50$ (or more) times smaller than the opposite case), although a more systematic dependence on the subsystem sizes is seen. The subsystems' dynamics in the latter case may thus be considered as \textit{effectively} Markovian. We have seen similar results for paramagnetic-to-paramagnetic and ferromagnetic-to-ferromagnetic quenches as well.

The lower row of Fig.\ref{fig:fig2} shows the dependence of this degree, for the case of a 4-spin subsystem (other subsystems show similar behaviour), on various values of longitudinal and transverse fields after quenching from the paramagnetic ground state. Here again, no discernible pattern is seen although a crude decrease of the degree is seen with increasing transverse fields.

\begin{figure}[tp]
	\centering
        \includegraphics[width=4.2cm,height=3cm]{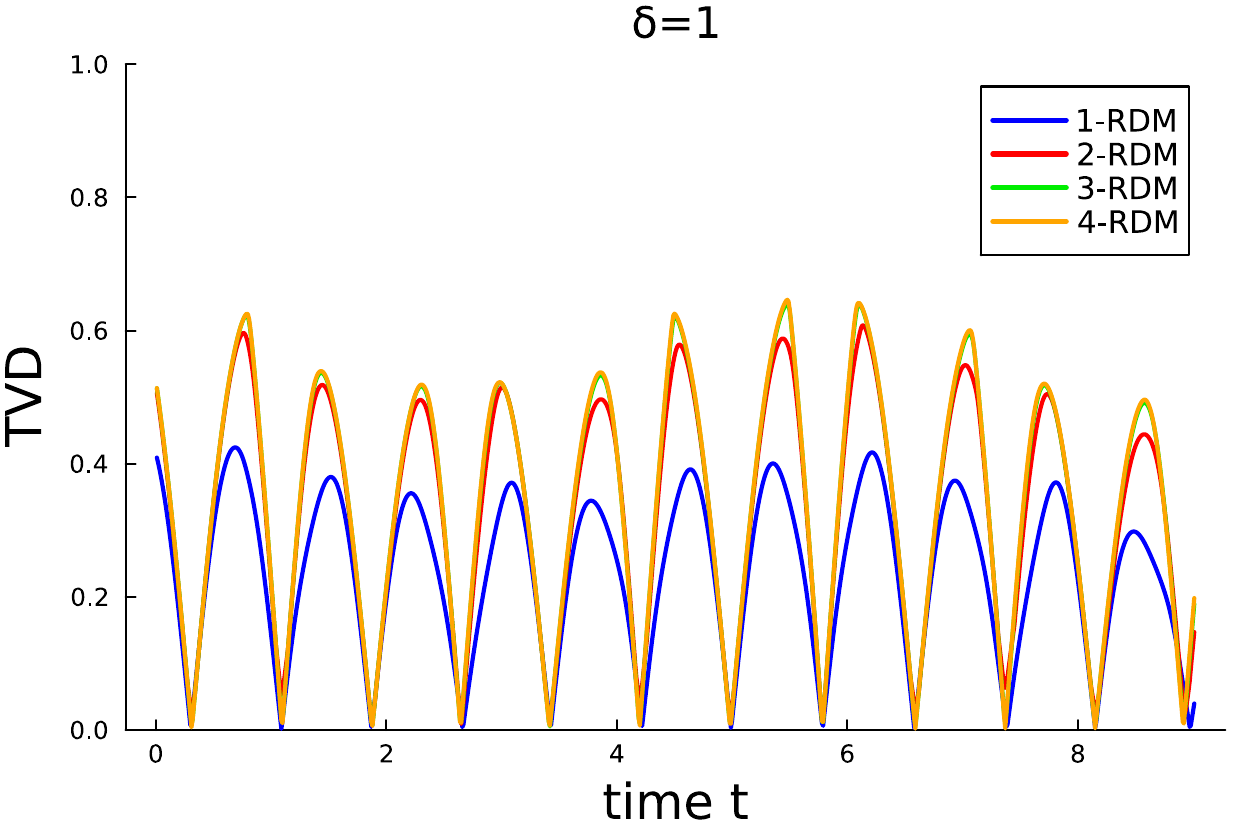}
        \includegraphics[width=4.2cm,height=3cm]{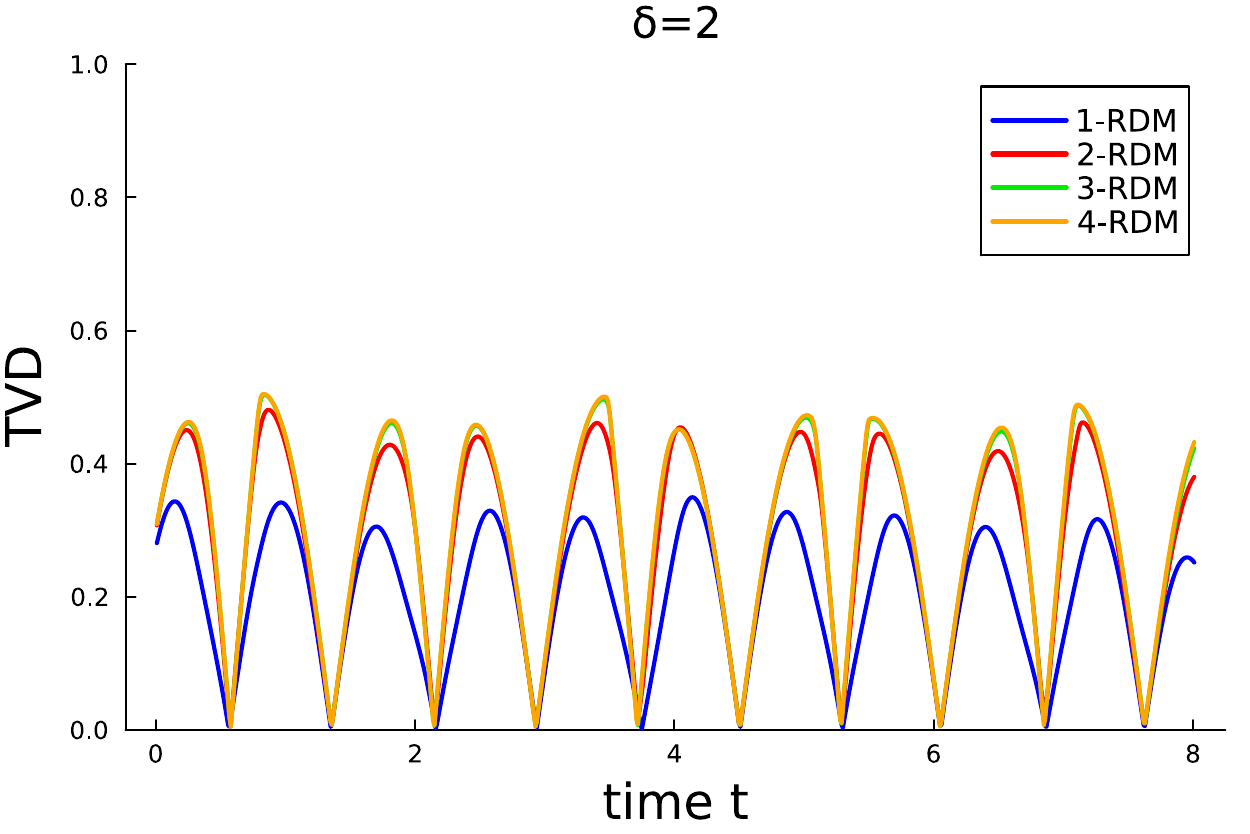}
        \caption{\fontsize{9}{11} \selectfont Strongly non-monotonic behaviour of total variation distances (TVD), Eq.(\ref{TVD}) (between descendingly-ordered eigenvalues of $\delta-$separated RDMs of small subsystems), on the y-axes vs time for $\delta\!=\!1$ (left) and $\delta\!=\!2$ (right) in the paramagnetic-to-ferromagnetic quench $(J,h_x,h_z) \!=\! (0.2,1,0) \rightarrow  (1,0.1,0.5)$. Here too, $k-$RDMs mean the same as in Fig.\ref{fig:fig1}. Note the existence of a timescale of $\sim 0.78 (\pm 0.02)$ in evolution times $t$ between the minima, which is largely independent of the subsystem size and of the temporal-separation $\delta$.}
        \label{fig:fig3}
\end{figure}

\begin{figure}[tp]
	\centering
	\includegraphics[width=4.2cm,height=3cm]{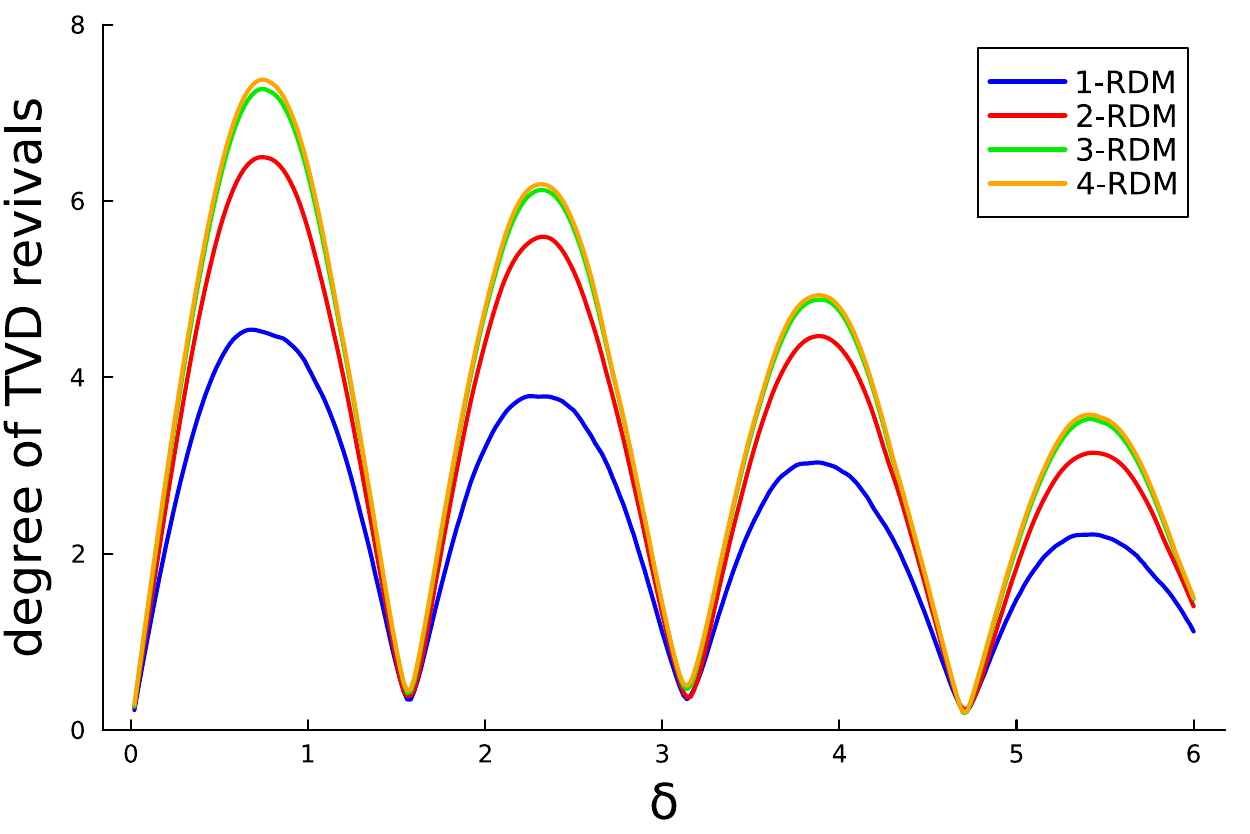}
        \includegraphics[width=4.2cm,height=3cm]{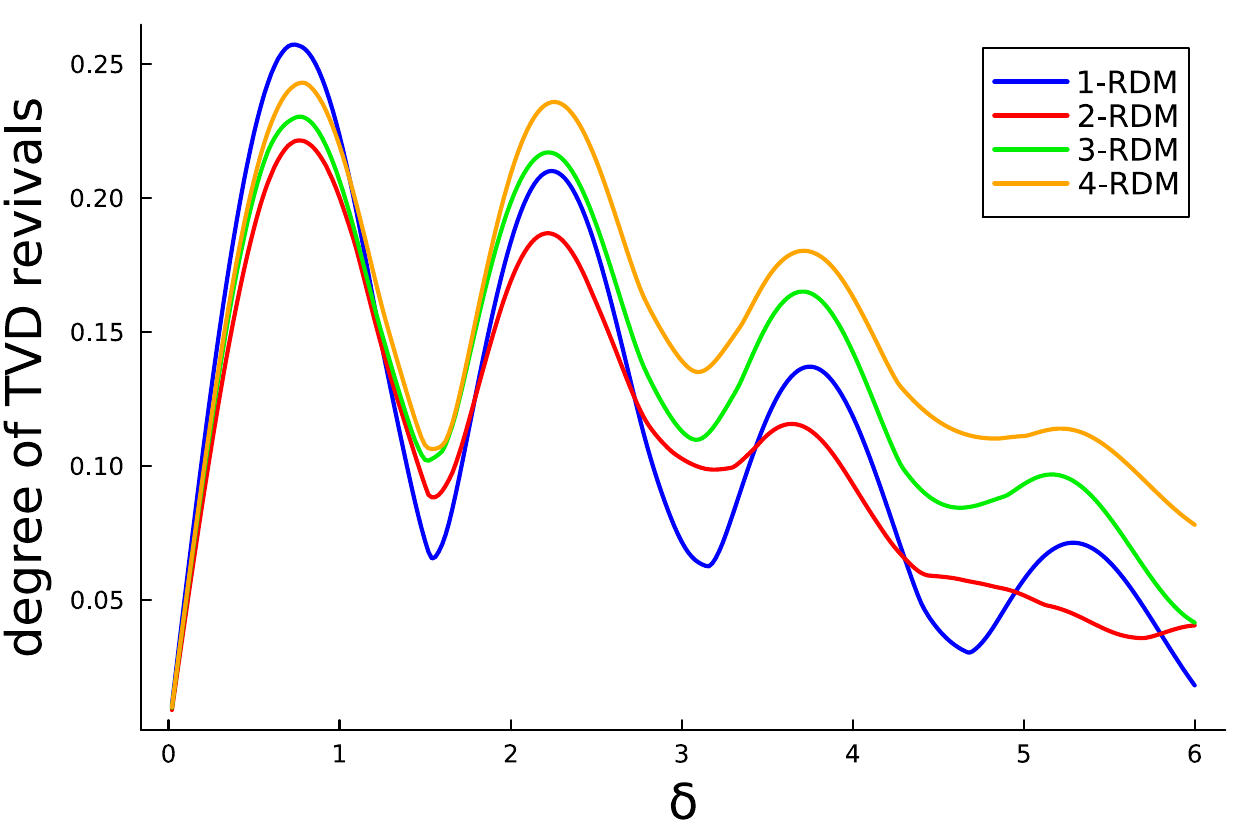}
        \includegraphics[width=4.2cm,height=3cm]{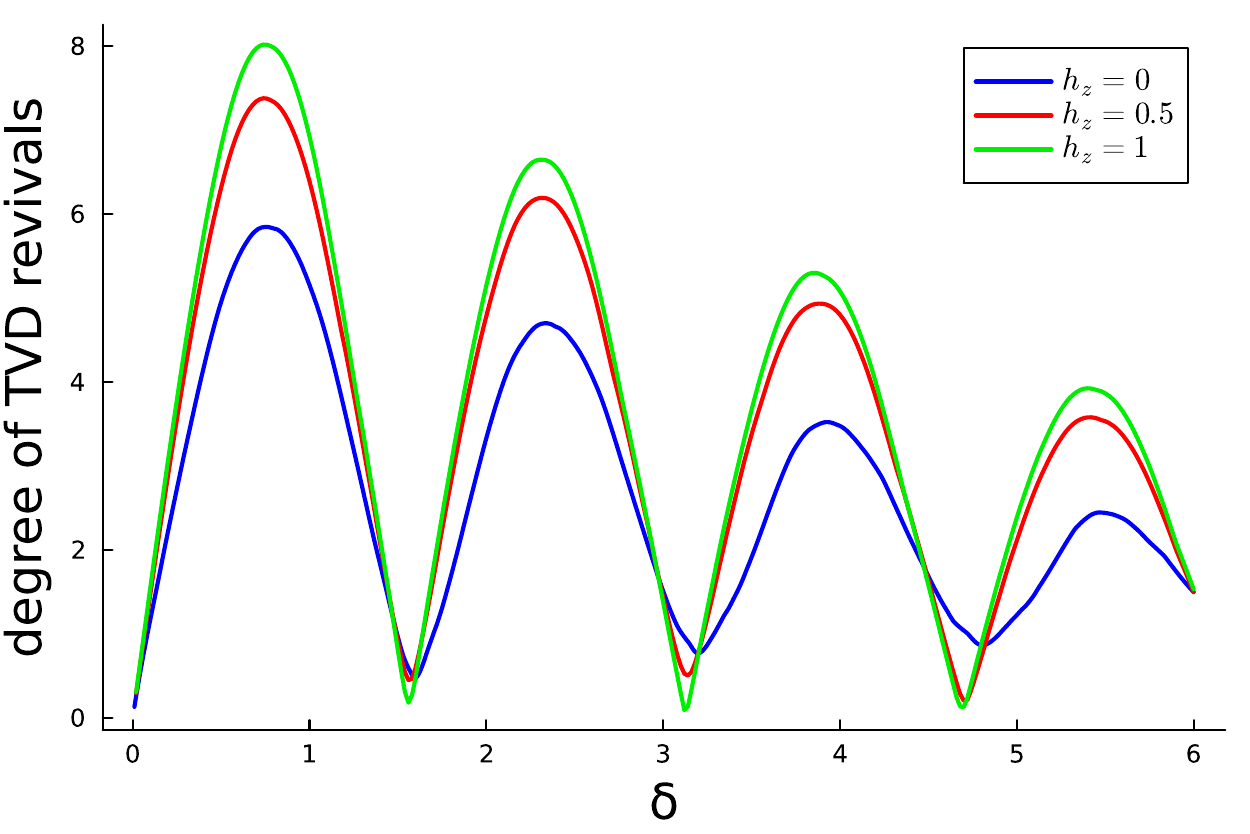}
        \includegraphics[width=4.2cm,height=3cm]{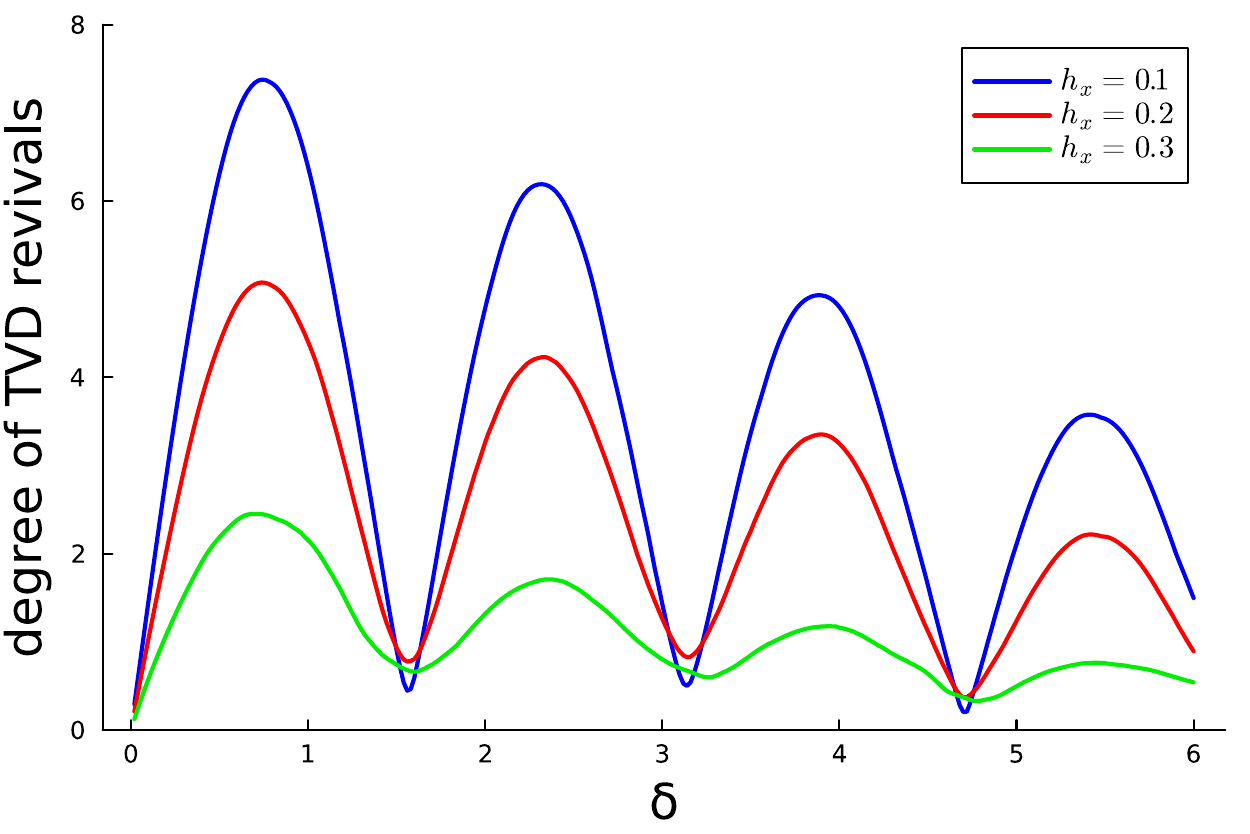}
        \caption{\fontsize{9}{11} \selectfont Degree of TVD revivals $\mathcal{D}_1(\delta)$, Eq.(\ref{TVDdegree}), on the y-axes for various subsystems vs $\delta$. \textbf{Upper Row} - For various small subsystems in paramagnetic to ferromagnetic quench $(J,h_x,h_z) \!=\! (0.2,1,0) \rightarrow  (1,0.1,0.5)$ (left), and ferromagnetic to paramagnetic quench $(J,h_x,h_z) \!=\! (1,0.1,0.5) \rightarrow  (0.2,1,0) $ (right). Here too, $k-$RDMs mean the same as in Fig.\ref{fig:fig1}. Note that the degree of TVD revivals in the latter quench are roughly 30 times smaller than in the former. Particularly prominent in the the former quench is the existence of a timescale of $\sim \{1.55,1.6\}$ in the temporal-separation $\delta$ between the maxima (or between the minima) in the degree of TVD revivals, which also is largely independent of the subsystem size. \textbf{Lower Row} - Dependence of $\mathcal{D}_1(\delta)$ on varying longitudinal field $h_z$ (left) and transverse field $h_x$ (right) for 4-RDMs in the former quench.}
        \label{fig:fig4}
\end{figure}

Remarkably, a more organized behaviour in the corresponding quantities is shown by the classical probability distributions furnished by the eigenvalues of the RDMs. Let us hypothesize for the moment that under CPTP maps acting on a pair of quantum states (density matrices), the distance between the corresponding probability vectors furnished by their respective eigenvalues also shows a monotonically non-increasing behaviour. Although intuitively appealing, we are not aware of any such results established in quantum information theory to the best of our knowledge. In classical information theory, distance measures between probability distributions also show monotonically non-increasing behaviour under classical processes and consequent inequalities are known as data-processing inequalities, see e.g. \cite{Cover1999}, whose quantum counterparts are the monotonic inequalities governing distances between quantum states under quantum channels/CPTP maps. But what classical "data processing" aspects are contained within quantum channels acting on quantum states is a subtle question in quantum information theory. If this hypothesis is true, could it be that any violation of this monotonicity under the action of a family of CPTP maps (dynamical maps) is also a signature of some aspect of non-Markovian dynamics of the underlying quantum states? It is intriguing to posit (and would be worthwhile to investigate systematically in future) that the quantum dynamics of subsystems also carry "classical" aspects in such a manner that the eigenvalues of respective reduced density matrices might show the classically-expected monotonicities for distance measures between them if the underlying quantum dynamics is Markovian, and a violation of those monotonicities when the underlying quantum dynamics is non-Markovian as per quantum measures. It is tempting to regard this as a classical characteristic of quantum non-Markovianity but in a sense different from \cite{Milz2020}.

Be that as it may, if the above hypothesis is true, then a degree of violation of said monotonicity may be defined again by the cumulative magnitude of the revivals (increases) of the employed distance measure. Fig.\ref{fig:fig3} shows highly non-monotonic behaviour of the total variation distances $V_d(q_{t+\delta}^{\ell}, q_t^{\ell})$ between descendingly-ordered sets of eigenvalues of RDMs (separated in time by $\delta\!=\!1,2$) corresponding to subsystems of size $\ell$, for the paramagnetic-to-ferromagnetic quench. In Fig.\ref{fig:fig4}, we show the corresponding degree of TVD revivals $\mathcal{D}_1(\delta)$ (Eq.\ref{TVDdegree}), which shows several remarkable facts : I) In the upper row, this degree for any considered subsystem is strongly oscillatory but decreasing on an average with increasing temporal separation $\delta$. II) The degrees for ferromagnetic-to-paramagnetic quench (right, upper row) are about $\sim 30$ times smaller than those for the paramagnetic-to-ferromagnetic quench (left, upper row). III) For paramagnetic-to-ferromagnetic quench (left, upper row), the $\delta$-values corresponding to the visible maxima and minima are nearly identical for all considered subsystems (notice however that the maxima corresponding to 1-RDMs are slightly shifted to the left relative to those corresponding to the other subsystems), and they appear to nearly coincide with each other already for subsystems as small as three and four spins. Note however that this degree at any given $\delta$ is larger in value for larger subsystems, in contrast to the degrees of non-Markovianity shown in Fig.\ref{fig:fig2}. Clearly then, this degree can not be a measure of non-Markovianity of quantum states in the usual sense. IV) In the lower row of Fig.\ref{fig:fig4} a general increase of these degrees with increasing $h_z$ is seen (for most values of $\delta$), whereas a rather drastic reduction in their oscillatory amplitude is seen with increasing $h_x$. Once again, quite remarkably the maxima and minima of these degrees at varying values of $h_z$ and $h_x$ occur at nearby $\delta$-values, with relative shifts becoming more prominent at higher $\delta$-values (especially for the integrable case $h_z=0$). 

Moreover, there seem to exist remarkable timescales associated with the behaviour of TVD between the eigenvalues of considered RDMs : a timescale of $\sim 0.78 (\pm 0.02)$ in evolution times $t$ between the minima in Fig.\ref{fig:fig3}, which is largely independent of the subsystem size and of the temporal-separation $\delta$ (we checked this for several $\delta$ values), and a second timescale of $\sim \{1.55,1.6\}$ in the temporal-separation $\delta$ between the maxima (or between the minima) emerges in the degrees of TVD revivals in Fig.\ref{fig:fig4} (upper row), which also is largely independent of the subsystem size (additionally, even with varying transverse and longitudinal fields there is a timescale of approximately the same value but in a cruder fashion, as seen in lower row of Fig.\ref{fig:fig4}).

\section{More discussion}      \label{sec5}

We now point to a heuristic argument for the pronounced non-Markovianity of subsystems' dynamics observed in the case of paramagnetic-to-ferromagnetic quench. In the computational basis of $\sigma_z$ spins with eigenstates denoted by $\mid\uparrow\rangle$ and $\mid\downarrow\rangle$, an ideal paramagnetic ground state of $N$ spins is a product state of the form $|+\rangle^{\otimes N}$, where at each site $|+\rangle \!=\! (\mid\uparrow\rangle + \mid\downarrow\rangle)/\sqrt{2}$, whereas the ideal ferromagnetic ground state (in the absence of fluctuations produced by the symmetry-breaking longitudinal field) takes the form of an $N$-partite Greenberger-Horne-Zeilinger (GHZ) state $(\mid\uparrow\rangle^{\otimes N}+\mid\downarrow\rangle^{\otimes N})/\sqrt{2}$, which is a highly entangled state. The former being a tensor product of equally weighted superpositions of the up and down spins at each site, while the latter being an equally weighted superposition of (tensor product of) up spins at all sites and down spins at all sites, it can be anticipated that transforming many-body states from the former to the latter is quite difficult to accomplish, and thus initiating a quench dynamics from the former to the latter as the targeted state in the long-time limit is a difficult demand to make from the system due to the very different microscopic structure of the initial and targeted states , and this plays a significant role in the considerable lack of a monotonous and memory-less relaxation dynamics. However, when the symmetry between up and down spins is broken by the presence of the longitudinal field (which additionally introduces confinement between kink-antikink excitations) and therefore the generalized GHZ state is no longer the target state, we have not seen any marked difference in the behaviour of the non-Markovianity degree with increasing longitudinal fields (however a distinctive behaviour was seen in the degree of TVD revivals), suggesting that in this case the emergence of confinement between the excitations replaces the former argument as the potentially underlying cause for the observed non-monotonic relaxation dynamics. The microscopic demands of the reversed quench dynamics from ferromagnetic to the magnetically disordered paramagnetic states is easier on the system as nearest-neighbor couplings (and thus couplings between subsystems and their environments) are weak, the excitations are now not subjected to confining forces nor the demands of any generalized GHZ-like target state, and a relatively quicker, monotonous and memory-less approach to equilibration follows.

We remarked earlier that the magnetically-ordered regime with non-zero longitudinal field is known to exhibit slow thermalization dynamics. This is mainly due to the existence of confining potential between certain excitations that separate domain walls \cite{McCoy1978,Delfino1996,Fonseca2003,Rutkevich2008}, as well as due to the presence of scars (non-thermalizing eigenstates) deep in the spectrum whose presence is related to and indeed enhanced by the presence of confinement \cite{James2019,Robinson2019}. The strength of the confining potential is directly proportional to the longitudinal field and we have seen in this article that, while the trace distance degrees signifying subsystem non-Markovianity did not show sufficiently differential behaviour with increasing values of the longitudinal field, the degrees of non-monotonicities of total variation distance between eigenvalues of subsystem RDMs showed marked increase with increasingly higher longitudinal fields. This intriguing connection would be worth understanding more precisely in future. Additionally, the asymmetry between the two directions of quenches, seen in this work in the context of subsystem non-Markovianity, was also noted previously in the context of dynamical quantum phase transitions and differentiating dynamical behaviour was seen in quantum mutual information (between small subsystems) and in leading members of the entanglement spectrum \cite{Nicola2021}, and many more quantum informationally distinguishing differences between the quench directions have been reported in \cite{Banerjee2025}. Both of these works have highlighted the driving role played by entanglement in the pronounced features shown in the paramagnetic-to-ferromagnetic quenches as opposed to the opposite quench.

 Despite the weak nearest-neighbor couplings in the paramagnetic side already earmarking subsystem dynamics to be within the Born-Markov paradigm, but noting that the interaction strengths between intra-subsystem spins, intra-environment spins and inter-subsystem-environment spins are also all equal (or comparable) to each other, with the environment also simultaneously being far from equilibrium, it may appear somewhat surprising that the ferromagnetic-to-paramagnetic quenches induce effectively Markovian dynamics of subsystems (at least as per the behaviour of distance measures between quantum states) and at least marginally noticeable (even if weakened) non-Markovianity may have been expected \textit{a priori}, since it seems more obvious to expect that subsystems of closed quantum many-body systems should naturally exhibit some notable degree of non-Markovianity in the far-from-equilibrium times for quenches between any parameter regimes because the process of equilibration of a closed quantum many body system is a complex process, with various subsystems interacting with (and acting as "environments" for) each other in a common pursuit to attain a globally equilibrated state. Therefore one might argue that the dynamics of a given subsystem (especially one that is sufficiently small compared to the total system size) plus its complement can not in general be expected to flow monotonously (i.e., a monotonous flow of information out of the subsystem to its environment) to equilibration in general, even more so because the environments themselves are also evolving dynamically. Yet, as we have seen it is the direction of quenches and the consequent restructuring and mixing of the quantum states that may be the most decisive factor on whether the ensuing dynamics of subsystems is Markovian or otherwise (in fact, general arguments suggest that subsystem dynamics being effectively Markovian should be more typical in most quantum systems' closed dynamics \cite{Romero2019}). These are crucially affected by the underlying physical scenarios as discussed in the previous paragraphs, such as the nature of entanglement creation needed to approach the target state, and the freeness or confinement of excitations to propagate between the subsystems in order to bring about the needed mixing and restructuring of the Hilbert space. Thus, the quasiparticle picture of quench dynamics \cite{Calabrese2005,Rieger2011,Alba2018}, where the elementary excitations propagate almost ballistically to bring about a linear growth of entanglement and fast equilibration dynamics, is the probable cause of the effectively Markovian dynamics of subsystems in the ferromagnetic-to-paramagnetic quenches. It is already known that this picture is violated in models with confinement, thus linking satisfaction or violation of this picture with consequent Markovianity or its lack thereof of the dynamics of subsystems of closed quantum many-body systems. 

 Furthermore, whenever subsystem dynamics is essentially Markovian, it is in principle plausible and would be worthwhile to derive Markovian master equation based descriptions of subsystems' dynamics and consequently develop an understanding of equilibration or thermalization and the ETH picture from this perspective. Recently, thermalization and ETH in open quantum systems have been studied in several contexts \cite{Moudgalya2019,Shirai2020,Ptaszynski2025,Purkayastha2024,Odonovan2024}, and it is hoped that similar descriptions could be extended to the understanding of approach to thermalization and ETH in closed quantum many-body systems from this perspective. A different approach to a subsystem based understanding of equilibration, thermalization and the emergence of statistical mechanics in closed quantum many-body systems is already developed in e.g. \cite{Tasaki1998,Goldstein2006,Linden2009,Short2011,Gemmer2009}.

\section{Conclusion}    \label{sec6}

By numerically investigating the dynamical behaviour of trace distances between temporally-separated reduced density matrices corresponding to small subsystems of a closed quantum spin system in one dimension, and with a viewpoint of looking at these subsystems as open quantum systems, we have presented numerical evidence of non-Markovianity of these subsystems' dynamics and the dependence of this characteristic on various factors at play. Notably, it was demonstrated that the direction (in parameter space) of quenching is a major factor for noticeable non-Markovianity, and that smaller subsystems showcase larger non-Markovianity in their dynamics in this quench dynamics (contrary to the results of \cite{Romero2019} that are more generally applicable, to which it seems the "extreme" paramagnetic-to-ferromagnetic quenches apparently constitute a (plausibly small) set of exceptions). We have also revealed a remarkably systematic dynamical behaviour in the measure of distance between the corresponding eigenvalues of the considered reduced density matrices, the interpretation and explanation of which we can not offer at this time and will be left for future work. We hope this sort of a viewpoint can be put to practice more widely by applying some of the central tools and concepts of open quantum systems theory \cite{Breuer2002,Vacchini2024} onto uncovering and understanding many facets of non-equilibrium dynamics of isolated quantum many-body systems (see also a recent work in this spirit \cite{Akhouri2025}). Specific to the issue of subsystem non-Markovianity described in this work, it would be worthwhile in future to derive an analytical description of these results perhaps along the lines of e.g. \cite{Znidaric2011,Apollaro2011,Garrido2016,Bhattacharya2017} by invoking the properties of generalized Bloch balls \cite{Kimura2003}.

While we focused in this article on the Ising spin model for simplicity and the paradigmatic value held by this model, we have observed similar signatures of subsystem non-Markovianity in quench dynamics of Heisenberg spin systems also, albeit they were most pronounced in extreme quenches from the paramagnetic to magnetically ordered phases in those systems as well, hence our focus on the simpler Ising spin chain in this article. In a recent work \cite{Banerjee2025}, we have revealed certain fine-grained attributes of entanglement dynamics for the same quench dynamics as in this work, and these go (possibly necessarily) hand-in-hand with subsystem non-Markovianity. Further important questions that need to be addressed include the effects of performing measurements on subsystem non-Markovianity, whether effective spin-boson type models or collision model descriptions can be constructed to mimic the open system dynamics of subsystems of closed many-body systems, a characterization of the properties of the dynamical maps themselves, and whether other approaches for characterizing and simulating non-Markovianity can provide additional or deeper insights. Work is in progress in some of these directions and we hope to be able to report on them in future.

\textbf{Acknowledgements---} We thank the Department of Atomic Energy, India for financial support.

%

\end{document}